\documentclass[12pt]{article}

\usepackage{bm}

\begin{document}

\title{Generalized Neutrino Equations}
\author{ Valeriy V. Dvoeglazov\\
UAF, Universidad Aut\'onoma de Zacatecas\\
Apartado Postal 636, Zacatecas 98061 Zac., M\'exico\\
E-mail: valeri@fisica.uaz.edu.mx\\
URL: http://fisica.uaz.edu.mx/\~{}valeri/\\
}

\date{\empty} 

\maketitle

%\medskip

\begin{abstract}
I discuss generalized spin-1/2 equations
for neutrinos. They have been obtained by means of the Gersten's method
for derivation of arbitrary-spin relativistic equations. Possible
physical consequences are discussed.
\end{abstract}

%\newpage

\section{Introduction}

Gersten~\cite{Ger} proposed a method for derivations
of massless equations of arbitrary-spin particles.
In fact, his method is related to the van der Waerden-Sakurai~\cite{Sak}
procedure for the derivation  of the massive Dirac equation. I commented
the derivation of Maxwell equations~[1a]\footnote{In fact, the $S=1$ quantum
equations.} in~\cite{dvo1}. Then, I showed that the method is rather ambigious
because instead of free-space Maxwell equations one can obtain {\it
generalized} $S=1$ equations, which connect the antisymmetric tensor field
with additional scalar fields. The problem of physical significance of
additional scalar chi-fields should be solved, of course, by experiment.

In the present article I apply the van der Waerden-Sakurai-Gersten
procedure to the spin-1/2 fields. As a result one obtains equations which
{\it generalize} the well-known Weyl equations. However, these equations
are known for a long time~\cite{antec}. 
Raspini~\cite{Rasp1,Rasp2,Rasp3,Rasp4,Rasp5} analized them again in detail.
I add some comments on  physical
contents of the generalized spin-1/2 equations.

\section{Derivation}

I use the equation (4) of the Gersten paper~[1a]
for the two-component spinor field function:
\begin{equation}
(E^2 -c^2 \vec{\bf p}^{\,2}) I^{(2)}\psi =
\left [E I^{(2)} - c \vec {\bf p}\cdot \vec{\bm \sigma} \right ]
\left [E I^{(2)} + c \vec {\bf p}\cdot \vec{\bm \sigma} \right ]
\psi = 0\qquad (eq. (4)\, of~[1a])\,.\label{G1}
\end{equation}
Actually, this equation is the massless limit of the equation which
has been  presented (together with the 
corresponding method of derivation of the Dirac equation)
in the Sakurai book~\cite{Sak}. In the latter case one
should substitute $m^2 c^4$ into the right-hand side of eq.~(\ref{G1}).
However, instead of equation (3.25)
of~\cite{Sak} one can define the two-component `right' field function
\begin{equation}
\phi_R= {1\over m_1 c} (i\hbar {\partial \over \partial x_0}
-i\hbar {\bm \sigma}\cdot {\bm \nabla}) \psi,\quad\phi_L=\psi\,
\end{equation}
with an additional mass parameter $m_1$.
In such a way we come to the system of the first-order differential equations
\begin{eqnarray}
&&(i\hbar {\partial \over \partial x_0}
+i\hbar {\bm \sigma}\cdot {\bm \nabla}) \phi_R
={m_2^2 c\over m_1}\phi_L\,,\\
&&(i\hbar {\partial \over \partial x_0}
-i\hbar {\bm \sigma}\cdot {\bm \nabla}) \phi_L
=m_1 c\phi_R\,.
\end{eqnarray}
It can be re-written
in the 4-component form:
\begin{eqnarray}
\label{gde}
&&\pmatrix{i\hbar (\partial/\partial x_0) &
i\hbar {\bm \sigma}\cdot {\bm \nabla}\cr
-i\hbar {\bm \sigma}\cdot {\bm \nabla}&
-i\hbar (\partial/\partial x_0)}
\pmatrix{\psi_A\cr\psi_B} =\\ 
&=&{c\over 2}
\pmatrix{(m_2^2/m_1
+m_1)&
(-m_2^2/m_1 +
m_1)\cr
(-m_2^2/m_1 +
m_1)& (m_2^2/m_1
+m_1)\cr}\pmatrix{\psi_A\cr\psi_B\cr}\nonumber
\end{eqnarray}
for the function $\Psi = column (\psi_A\quad \psi_B)=
column (\phi_R+\phi_L\quad \phi_R - \phi_L )$.
The equation (\ref{gde}) can be written in the covariant form.
\begin{equation}
\left [ i\gamma^\mu \partial_\mu - {m_2^2 c\over m_1 \hbar}
{(1-\gamma^5)\over 2} -{m_1 c \over \hbar} {(1+\gamma^5)\over 2}
\right ]\Psi = 0\,.\label{gde1}
\end{equation}
The standard representation of $\gamma^\mu$ matrices
has been used here.

If $m_1=m_2$ we can recover the standard Dirac equation.
As noted in~[4b] this procedure can be viewed as the simple change of
the representation of $\gamma^\mu$ matrices (unless
$m_2 \neq 0$).

Furthermore, one can  either repeat a similar procedure
(the modified Sakurai procedure) starting from the {\it massless}
equation (4) of~[1a] or put $m_2=0$ in eq.  (\ref{gde1}). The {\it massless
equation} is\footnote{It is necesary to stress that the term {\it `massless'}
is used in the sense that $p_\mu p^\mu =0$.}
\begin{equation}
\left [
i\gamma^\mu \partial_\mu - {m_1 c \over \hbar} {(1+\gamma^5)\over 2}
\right ]\Psi = 0\,.\label{gd1}
\end{equation}
Then, we may have different physical consequences following from (\ref{gd1}) with 
 those which follow from 
the Weyl equation.\footnote{Remember 
that the Weyl equation
is obtained as $m\rightarrow 0$ limit of the usual Dirac equation.} The
mathematical reason of such a possibility of  different massless
limits is that the corresponding change of representation of $\gamma^\mu$
matrices involves mass parameters $m_1$ and $m_2$ themselves. The corresponding 
transformation matrix 
may be non-existent (its elements
tend to infinity in the certain limit).

It is interesting to note that we can also repeat this procedure
for the definition (or for even more general definitions);
\begin{equation}
\phi_L= {1\over m_3 c} (i\hbar {\partial \over \partial x_0}
+i\hbar {\bm \sigma}\cdot {\bm \nabla}) \psi,\quad\phi_R=\psi\,.
\end{equation}
This is due to the fact that the parity properties of
the two-component spinor are undefined in the two-component equation.
The resulting equation is
\begin{equation}
\left [ i\gamma^\mu \partial_\mu - {m_4^2 c\over m_3 \hbar}
{(1+\gamma^5)\over 2} -{m_3 c \over \hbar} {(1-\gamma^5)\over 2}
\right ]\tilde\Psi = 0\,,
\end{equation}
which gives us  yet another equation in the massless limit ($m_4
\rightarrow 0$):
\begin{equation} \left [ i\gamma^\mu \partial_\mu - {m_3
c \over \hbar} {(1-\gamma^5)\over 2} \right ]\tilde\Psi = 0\,, \label{gd2}
\end{equation}

The above procedure can be generalized to {\it any} Lorentz group
representations, {\it i.~e.}, to any spins.
In some sense the equations (\ref{gd1},\ref{gd2}) are
analogous to the  $S=1$ equations~\cite[(4-7,10-13)]{dvo1}, which
also contain additional parameters.

\section{Physical Interpretations and Conclusions}

Is the physical content of the generalized $S=1/2$
{\it massless} equations the same as that of the Weyl equation?
Our answer is `No'. The excellent discussion can be found
in~[4a,b]. First of all, the theory does {\it not} have chiral invariance. Those authors
call the additional parameters as measures of the degree of chirality.
Apart of this, Tokuoka introduced the concept of the gauge transformations
(not to confuse with phase transformations) for the 4-spinor fields. He
also found some strange properties of the  anti-commutation relations
(see \S 3 in~[4a] and cf.~[11b]). And finally, the equation (\ref{gd1}) describes
{\it four} states, two of which answer for the positive energy $E=\vert
{\bf p}\vert$, and two others answer for the negative energy $E=-\vert
{\bf p}\vert$.

I just want to add the following to the discussion.
The operator of the {\it chiral-helicity} $\hat\eta = ({\bm\alpha}\cdot
\hat{\bf p})$ (in the spinorial representation) used in~[4b] (and
re-discovered in~[11a]) does {\it not} commute, {\it e.g.}, with the Hamiltonian of the
equation~(\ref{gd1}):\footnote{Do not confuse with the Dirac Hamiltonian.}
\begin{equation} [{\cal H}, {\bm\alpha}\cdot \hat{\bf
p} ]_- = 2{m_1 c \over \hbar} {1-\gamma^5 \over 2} ({\bm \gamma}\cdot
\hat{\bf p})\, .
\end{equation}
For the eigenstates of the {\it chiral-helicity} the system of corresponding
equations can be read ($\eta=\uparrow, \downarrow$)
\begin{equation} i\gamma^\mu
\partial_\mu \Psi_\eta - {m_1 c\over \hbar}{1+\gamma^5 \over 2}\Psi_{-\eta}
=0 \, .  \end{equation} The conjugated eigenstates of the Hamiltonian
$\vert \Psi_\uparrow + \Psi_\downarrow >$ and
$\vert \Psi_\uparrow - \Psi_\downarrow >$
are connected, in fact, by $\gamma^5$ transformation
$\Psi \rightarrow \gamma^5 \Psi \sim ({\bm\alpha}\cdot \hat{\bf p})
\Psi$ (or $m_1\rightarrow -m_1$).  However, the $\gamma^5$ transformation
is related to the $PT$ ($t\rightarrow - t$ only) transformation~[4b],
which, in its turn, can be interpreted as $E\rightarrow -E$, if one
accepts the Stueckelberg idea about antiparticles.  We associate $\vert
\Psi_\uparrow + \Psi_\downarrow >$ with the positive-energy eigenvalue of
the Hamiltonian $E=\vert {\bf p}\vert$ and $\vert \Psi_\uparrow -
\Psi_\downarrow >$, with the negative-energy eigenvalue of the
Hamiltonian ($E=-\vert{\bf p}\vert$). Thus, the free chiral-helicity
massless eigenstates may oscillate one to another with the frequency
$\omega = E/\hbar$ (as the massive chiral-helicity eigenstates, see~[10a]
for details). Moreover, a special kind of interaction which is not
symmetric with respect to the chiral-helicity states (for instance, if
the left chiral-helicity eigenstates interact with the matter only) may induce
changes in the oscillation frequency, like in the Wolfenstein (MSW) formalism.

The question is: how can these frameworks be connected
with the Ryder method of derivation of relativistic wave equations, and
with the subsequent analysis of problems of the choice of
normalization and of the choice of  phase factors in the papers~\cite{Dvo2,Ahlu,Dvo3}?
However, the conclusion may be similar to that which was  achieved before: the
dynamical properties of the massless particles ({\it e.~g.}, neutrinos and photons)
may differ from those defined by the well-known Weyl and Maxwell equations.

{\bf Acknowledgments.}
I greatly appreciate old discussions with Prof. A. Raspini
and useful information from Prof. A. F. Pashkov. 
This work has been partly supported by
the ESDEPED.

\end{document}